\documentclass[10pt,showpacs,showkeys]{revtex4}

\usepackage{natbib}

 \usepackage{graphics}
 \usepackage{graphicx}
\usepackage{epsfig}
\usepackage{bm}

\usepackage{amssymb}

\begin{document}



\title{Quasi-One-Dimensional Quantum Ferrimagnets}


\author{R. R. Montenegro-Filho}
\author{M. D. Coutinho-Filho}
\affiliation{Laborat\'orio de F\'{\i}sica Te\'orica e Computacional, Departamento de F\'{\i}sica,
Universidade Federal de Pernambuco, 50670-901, Recife, PE, Brasil}

\begin{abstract}
We present an exact diagonalization study of the half-filled Hubbard model
on bipartite quasi-one-dimensional lattices.
In particular, we emphasize the dependence of the ferrimagnetic ground state properties,
 and its associated magnetic excitations, on the Coulomb repulsion $U$.

\end{abstract}



\maketitle


\section{Introduction}
In the last years the physics of quasi-one-dimensional compounds has been the
object of intense theoretical and experimental investigations.
In this work we study the ground state (GS) and magnetic excitations of
two bipartite chains motivated by low-dimensional inorganic \cite{hoffman}
and organic \cite{korshak} ferrimagnets:
the $AB_2$ chain in Fig. \ref{fsmhei}(a) and the $ABC$ chain in Fig. \ref{fsmhei}(b),
respectively.
We shall use the simplest approach for interacting electron systems
on a lattice with $N$ sites, namely the one-band Hubbard Model (HM):
\begin{equation}
\mathcal{H}=-t\sum_{ \langle i,j\rangle,\sigma}c_{i\sigma}^{\dagger}c_{j\sigma}+U\sum_{i}n_{i\uparrow}n_{i\downarrow},\label{hub}
\end{equation}
where $c_{i\sigma}^{\dagger}$ ($c_{i\sigma}$) is the creation (annihilation) operator
for electrons
with spin $\sigma$ at site $i$; $t$ is the hopping parameter and $U$ is the on site
Coulomb repulsion.
For the extremely localized regime ($U \gg t$) and $n\equiv N_{e}/N=1$,
where $N_e$ is the number of particles, the HM
can be mapped onto the Antiferromagnetic Heisenberg Model (AHM) \cite{raposo1}:
\begin{equation}
\mathcal{H}=\frac{J}{2}\sum_{ \langle i,j\rangle} \vec{S}_{i}\cdot\vec{S}_{j},\label{hei}
\end{equation}
where $J=4t^{2}/U$.
We should mention that Mac\^edo {et al.} \cite{mauricio}, using
a variety of numerical methods, have derived the ferrimagnetic
nature of $AB_2$ chains modeled
by the HM in the strong and weak coupling regimes. Here we focus on the
half-filled band case ($n=1$) and extend these previous studies to much
larger systems using Lanczos exact diagonalization technique.
\par It is also worth mentioning that the two lattices present
three electronic bands in the tight-binding limit ($U=0$): one
flat band at $\epsilon=0$ for the two chains;
and two dispersive ones, $\epsilon(q)=\pm 2t\sqrt{2}\cos(q/2)$ and $
\epsilon(q)=\pm t\sqrt{1+[2\cos(q/2)]^2}$,
with $q=2\pi l/(N/3)$ and $l=0...(N/3)-1$ for the
$AB_2$ and $ABC$ chains, respectively.
The flat band is closely associated with
ferrimagnetic properties of both chains at half-filling \cite{mauricio,flatband}.
A particular feature of the $AB_2$ chain is a local
invariance under the exchange of the $B$ sites in
any cell $l$ \cite{alcaraz}. The eigenvalues of the exchange
operator being $p_l=\pm 1$.
This symmetry leads to a
conserved local spatial parity that in the Heisenberg
limit asserts that the $B$ sites form either
a singlet ($p_l=1$) or a triplet ($p_l=-1$) bond state.
Therefore for the symmetry sector in which all $p_l=-1$
the spectrum is identical to that of the
alternating spin-$\frac{1}{2}$/spin-$1$ AHM chain \cite{alcaraz}.

\section{Ferrimagnetic Ground State}

\par
\begin{figure}[th]
\begin{center}
\centerline{\includegraphics*[width=66mm,clip]{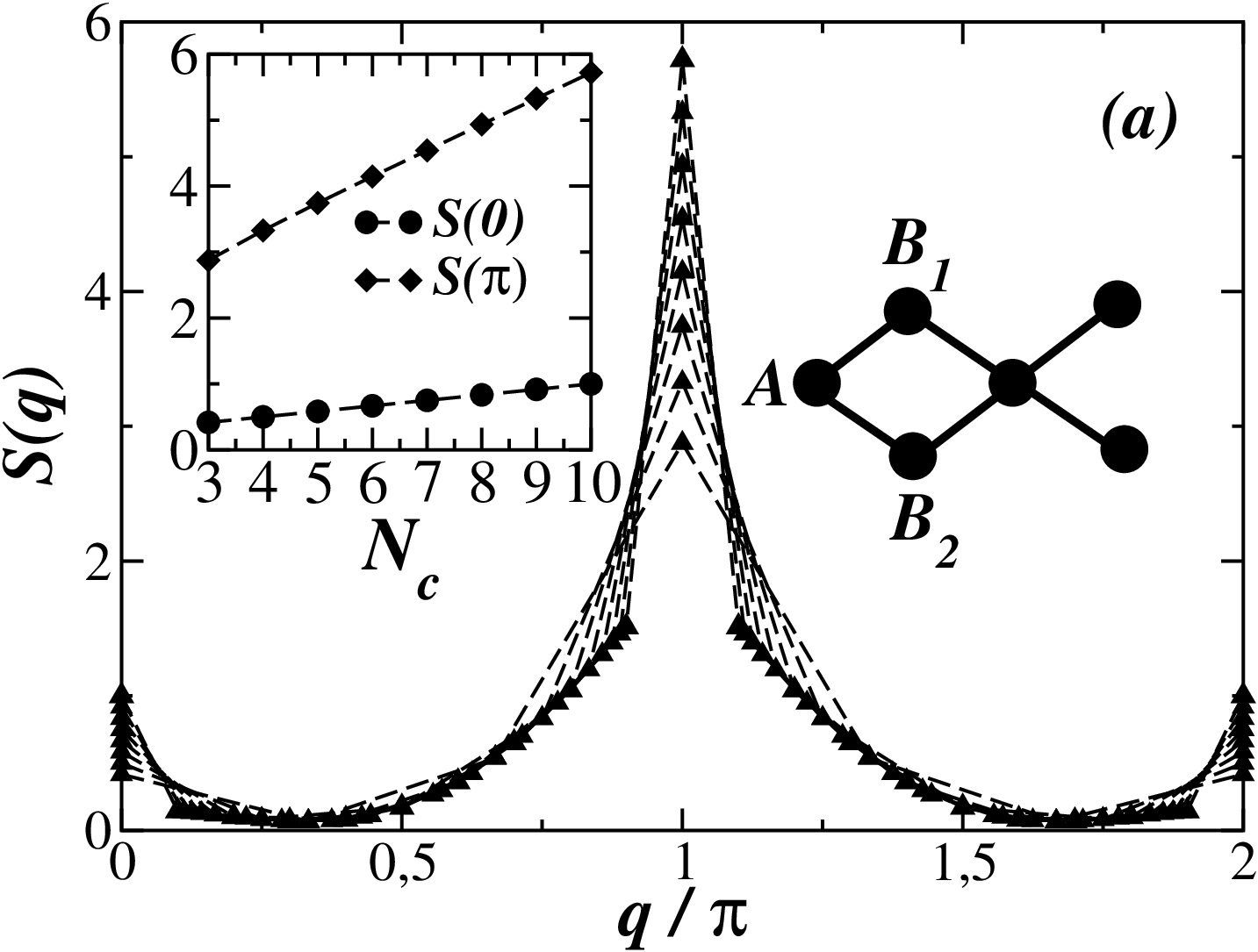}
\includegraphics*[width=64mm,clip]{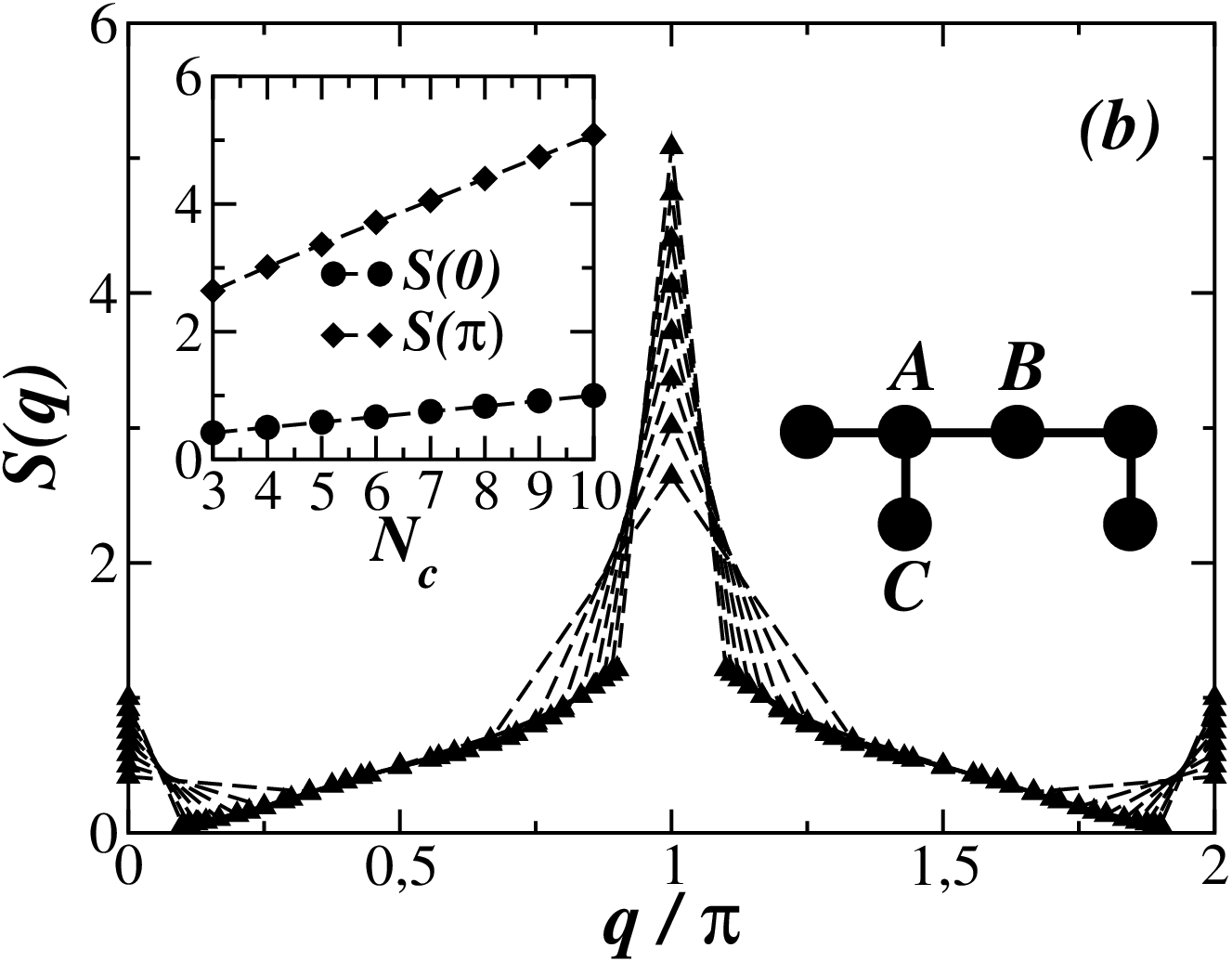}}
\caption{Magnetic Structure Factor $S(q)$ for the $AB_{2}$ (a) and $ABC$ (b) chains
in the Heisenberg limit ($U \gg t$). The size ranges from $N_c=3$ to $N_c=10$. The insets
display the size dependence of the Ferromagnetic [$S(0)$] and Antiferromagnetic
[$S(\pi)$] peaks. Dashed lines are guides for the eye.}
\label{fsmhei}
\end{center}
\end{figure}

 A theorem due to Lieb and Mattis \cite{lieb2} asserts that the
GS of the AHM on a bipartite lattice has total spin $S_{g}=|N_{1}-N_{2}|/2$,
where $N_{1}$ and $N_{2}$ are the number of lattice sites at the sub-lattices $1$ and
$2$, respectively. So, if $|N_{1}-N_{2}|\sim N$ the system display unsaturated
ferromagnetic (F) GS. The coexistence of both F
and antiferromagnetic (AF) long-range order implies in ferrimagnetism,
as rigorously proved by Tian \cite{tian1}. Another crucial step
was provided by Lieb \cite{lieb1}, who proved that the GS
of the HM for $n=1$ has $S_{g}=|N_{1}-N_{2}|/2$ for any $U>0$.
In this case,
Tian and collaborators \cite{tian2} further
established the ferrimagnetic (FERRI) long-range order of the GS if $|N_{1}-N_{2}|\sim N$.
The unit cell of the two chains of interest
has $2$ sites in one sublattice and $1$ in the other (see Fig. \ref{fsmhei}),
so that $S_g=N_c/2$, where $N_c=N/3$ is the number
of unit cells. However, although ferrimagnetism is expected, the specific magnetic structure
of a unit cell may strongly depend on the Coulomb coupling, as well as on quantum fluctuations,
and is not known {\it a priori} from the more general results of the theorems stated above.
\par
\begin{figure}[th]
\begin{center}
\centerline{\includegraphics*[width=65mm,clip]{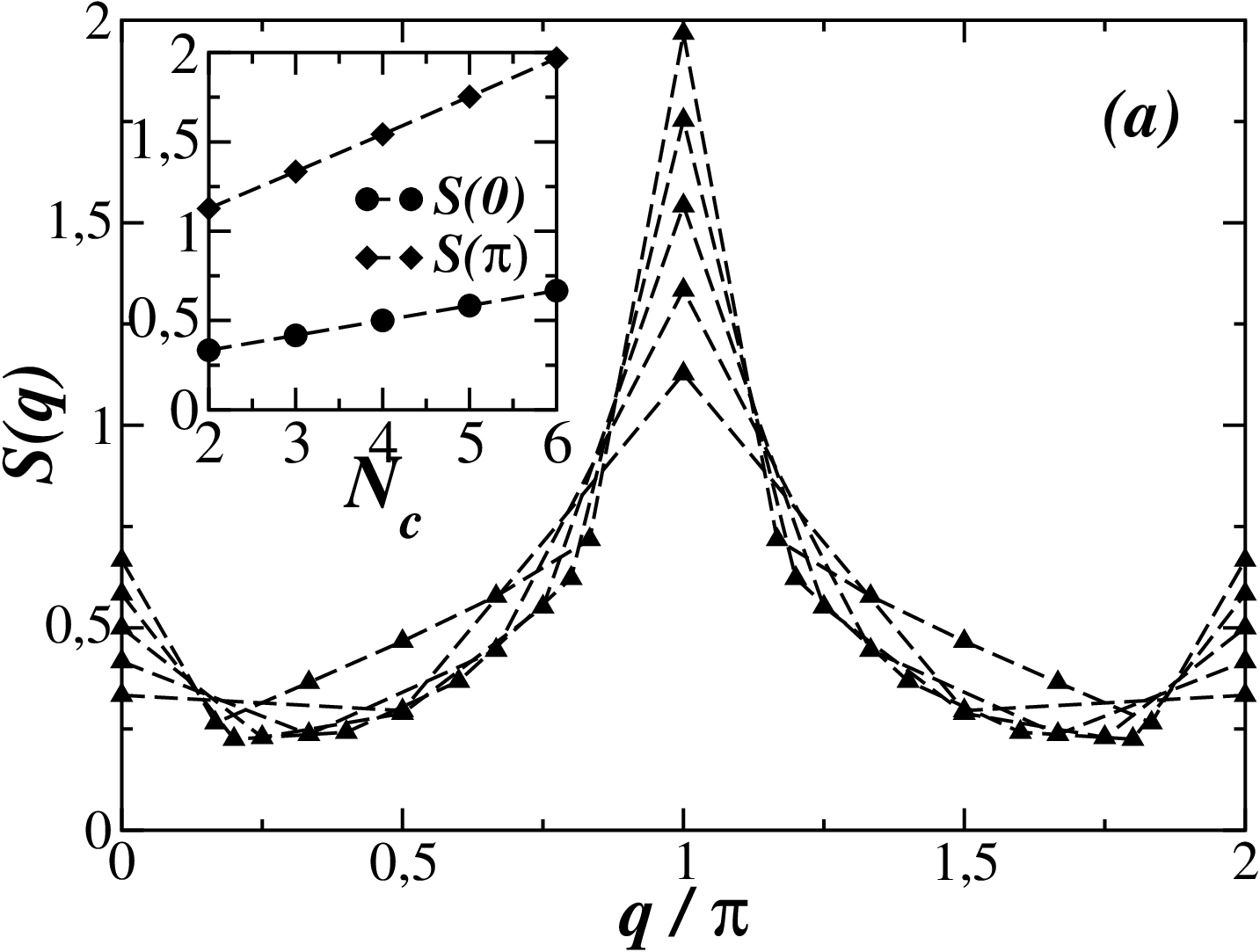}
\includegraphics*[width=65mm,clip]{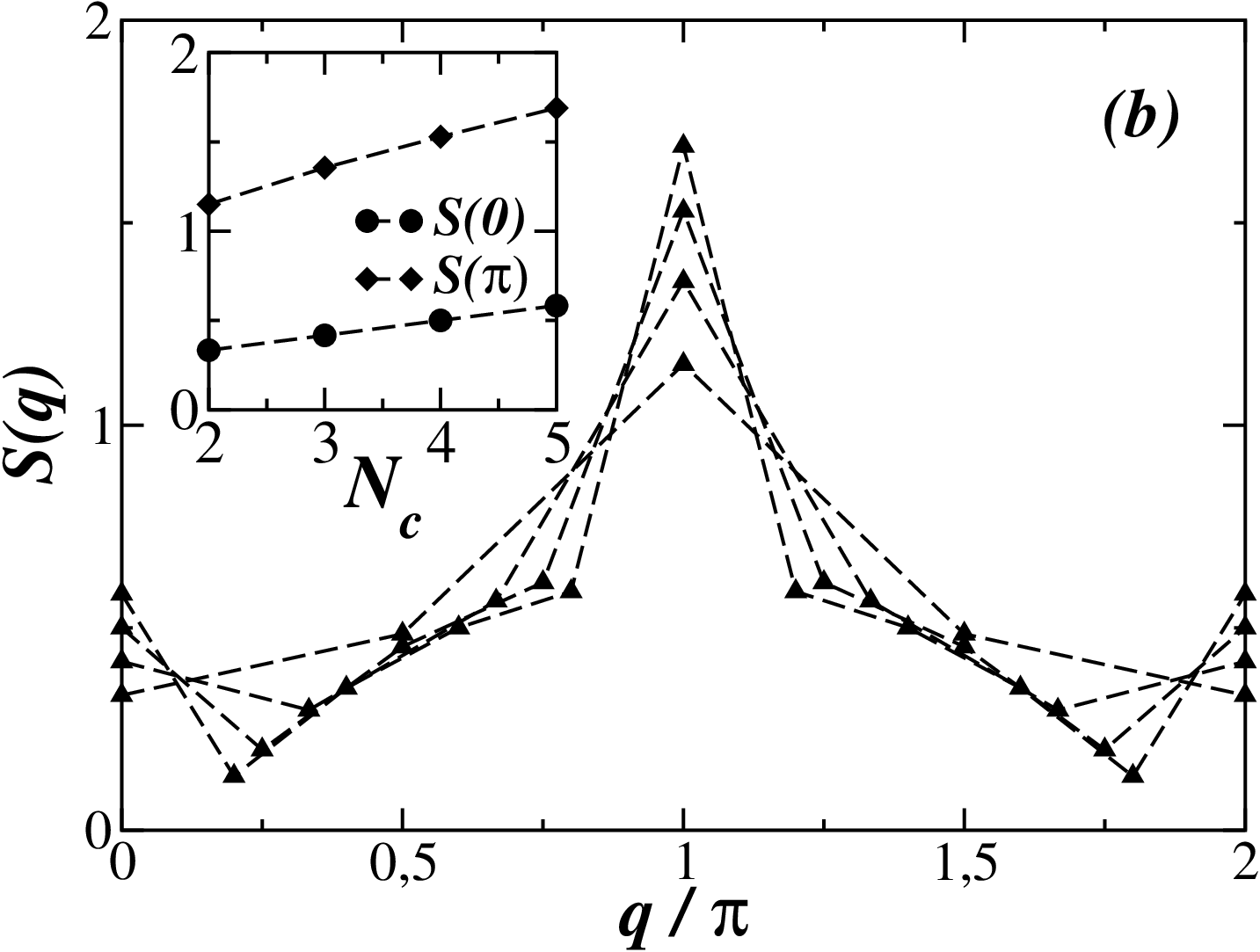}}
\caption{Magnetic Structure Factor $S(q)$ for the $AB_{2}$ (a) and $ABC$ (b) chains
using the HM for $U=2t$. The size ranges from $N_c=2$ to $N_c=6$ for the $AB_2$ chain 
and to $N_c=5$ for the $ABC$. The inset
presents the size dependence of the Ferromagnetic [$S(0)$] and Antiferromagnetic
[$S(\pi)$] peaks. Dashed lines are guides for the eye.}
\label{fsmhub}
\end{center}
\end{figure}

 We can probe the
magnetic order through the Magnetic Structure Factor (MSF):
\begin{equation}
S(q)=\frac{1}{N}\sum_{i,j} e^{iq(x_{i}-x_{j})} \langle \vec{S}_{i}\cdot\vec{S}_{j}\rangle,
\end{equation}
which is related to the zero-field static magnetic susceptibility by $\chi(q=0)=S(0)/(k_B T)$,
where $k_B$ is the Boltzmann constant and $T$ is the temperature.
The condition for a F (AF) ordered state is that $S(0) [S(\pi)]\sim N$,
so that in a long-range FERRI GS the two conditions must be fulfilled.
This is indeed the case for the two chains, as shown in Figs. \ref{fsmhei} and \ref{fsmhub}, both
in the strong and weak coupling limits.

\par Due to the critical nature of both chains at low temperatures,
the correlation length $\xi$ and $\chi(q=0)$ satisfy power law behaviour:
$\xi\sim T^{-\nu}$ and $\chi\sim T^{-\gamma}$ as $T\rightarrow 0$.
Since $\xi \sim N$ at $T=0$, using scaling arguments and the results of Fig. \ref{fsmhei},
we have $T^{-\gamma}\sim T^{-\nu}/T$, i. e., $\gamma-\nu=1$, in agreement with
the values $\gamma=2$ and $\nu=1$ derived using renormalization group
techniques \cite{raposo2}.
\par
\begin{figure}[th]
\begin{center}
\centerline{\includegraphics*[width=66mm,clip]{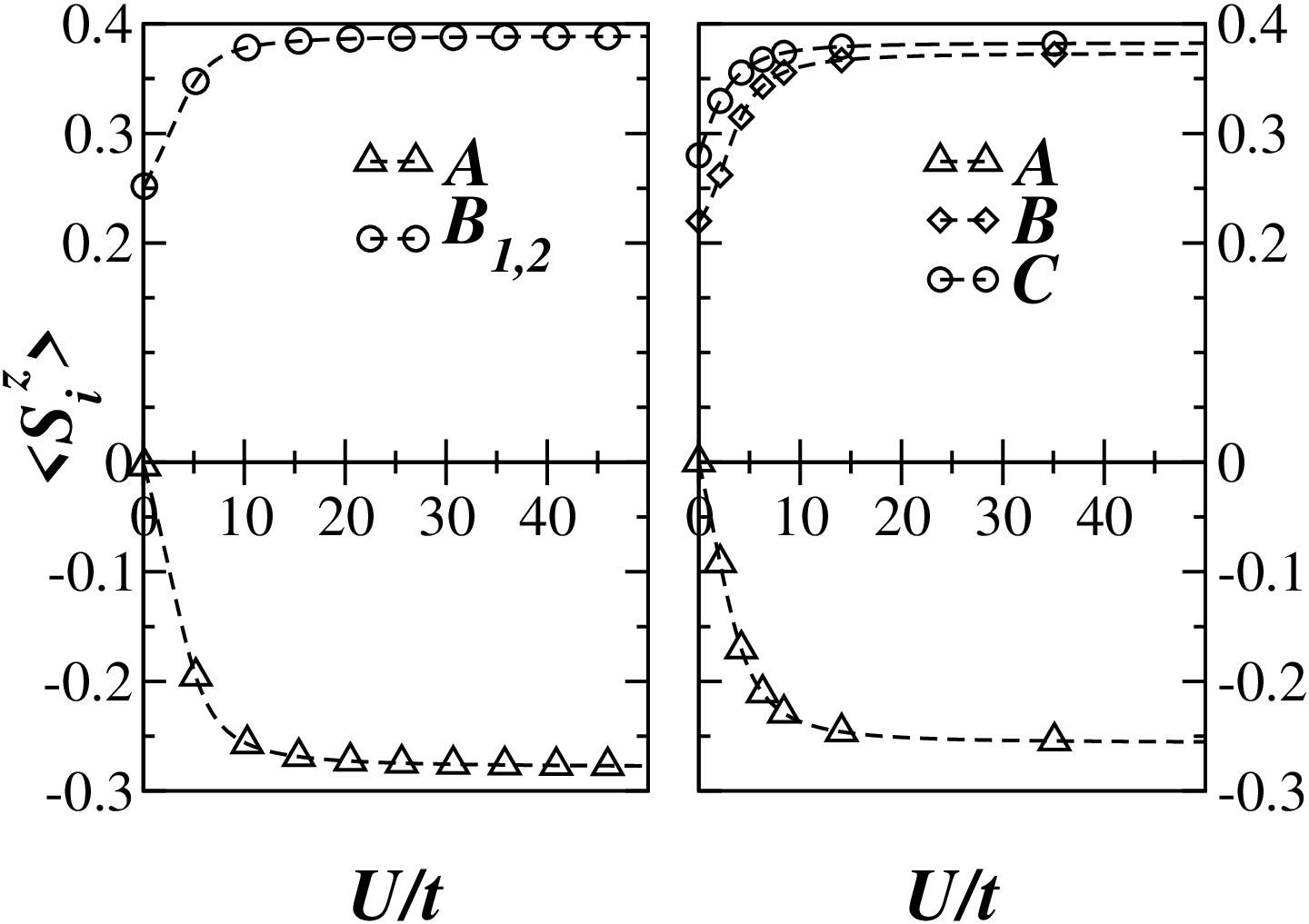}
\includegraphics*[width=65mm,clip]{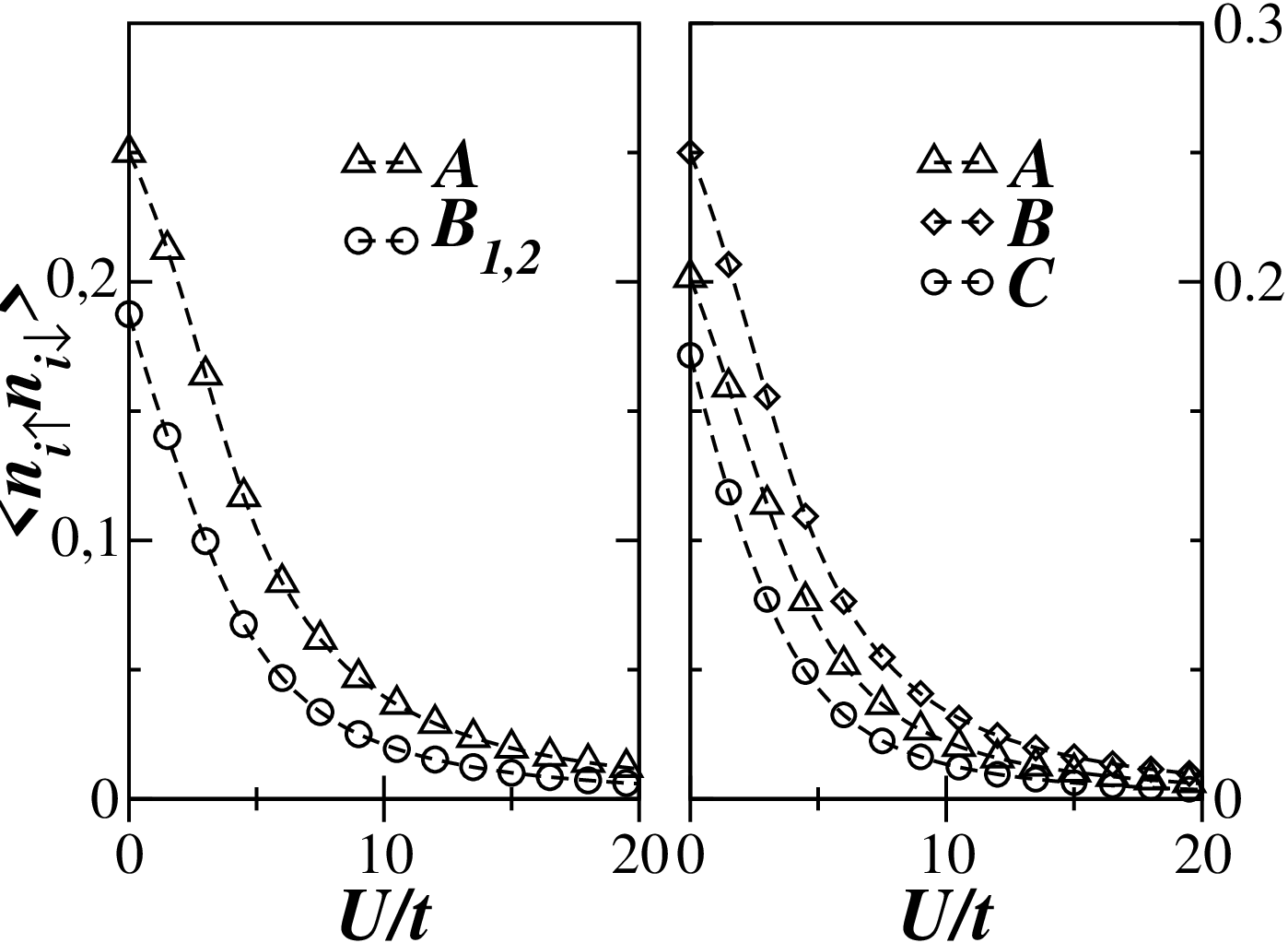}}
\caption{Site magnetization $ \langle S^{z}_i\rangle$ and average site
double occupancy $ \langle n_{i\uparrow}n_{i\downarrow}\rangle$
as function of $U$ for
$N_c=5$ in the $AB_{2}$ and $ABC$ chains in the spin sector
$S^{z}=S_{g}$. Dashed lines are guides for the eye.}
\label{smfu}
\end{center}
\end{figure}

 In Fig. \ref{smfu} we present the local magnetization ($S^{z}_{i}$)
and the average local value of double occupancy ($ \langle n_{i\uparrow}n_{i\downarrow}\rangle$)
as function of $U$. Since the system has a spontaneously broken
$SU(2)$ symmetry, we can choose the spin sector $S^{z}=S_{g}$ to study
these quantities, where $S^{z}$ is the total spin $z$ component.
In the Heisenberg limit ($U \gg t$, $ \langle n_{i\uparrow}n_{i\downarrow}\rangle$=0), the
bulk magnetization
estimated from chains up to $N_c=10$ are: $ \langle S^{z}_{A}\rangle=-0.2925$ and
$ \langle S^{z}_{B_{1,2}}\rangle=0.3962$,
for the $AB_{2}$ chain, in agreement with density matrix renormalization group (DMRG)
calculations \cite{white} and with values for $ \langle S^{z}_{A}\rangle$ and $2 \langle S^{z}_{B_{1,2}}\rangle$
in the spin-$\frac{1}{2}$/spin-$1$
chain \cite{spin1spin2}.
For the $ABC$ chain, $ \langle S^{z}_{A}\rangle=-0.2567$, $ \langle S^{z}_{B}\rangle=0.3741$
and $ \langle S^{z}_{C}\rangle=0.3826$.
Notice that a FERRI structure sustain quantum fluctuations, which strongly reduce
the local magnetization from the classical value,
although in any case the unit cell magnetization remains $1/2$.
In the non-interacting limit ($U=0$) we see in Figs. \ref{smfu}(a) and \ref{smfu}(b)  that
$ \langle S^{z}_{A}\rangle=0$ and $\langle S^{z}_{B_1(B)}+S^{z}_{B_2(C)}\rangle =0.5$ for
the $AB_2$ ($ABC$) chain. These results being related to the band structure:
filling the bands up to $n=1$ and choosing the spin sector $S^{z}=S_{g}$,
there exist $N_c$ unpaired electrons in the flat band which occupy
 $B_1(B)$ and $B_2(C)$ sites only.
It is important to note that in this limit the global GS does not have an unique
value of $S$, but rather $S\leq N_c/2$. As the Coulomb coupling is turned on, the GS, with a unique
$S_g=N_c/2$, evolves adiabatically
from the weak to the strong coupling regime (Fig. \ref{smfu}).
\section{Magnetic Excitations}
\par
\begin{figure}[th]
\begin{center}
\centerline{\includegraphics*[width=6.5cm,clip]{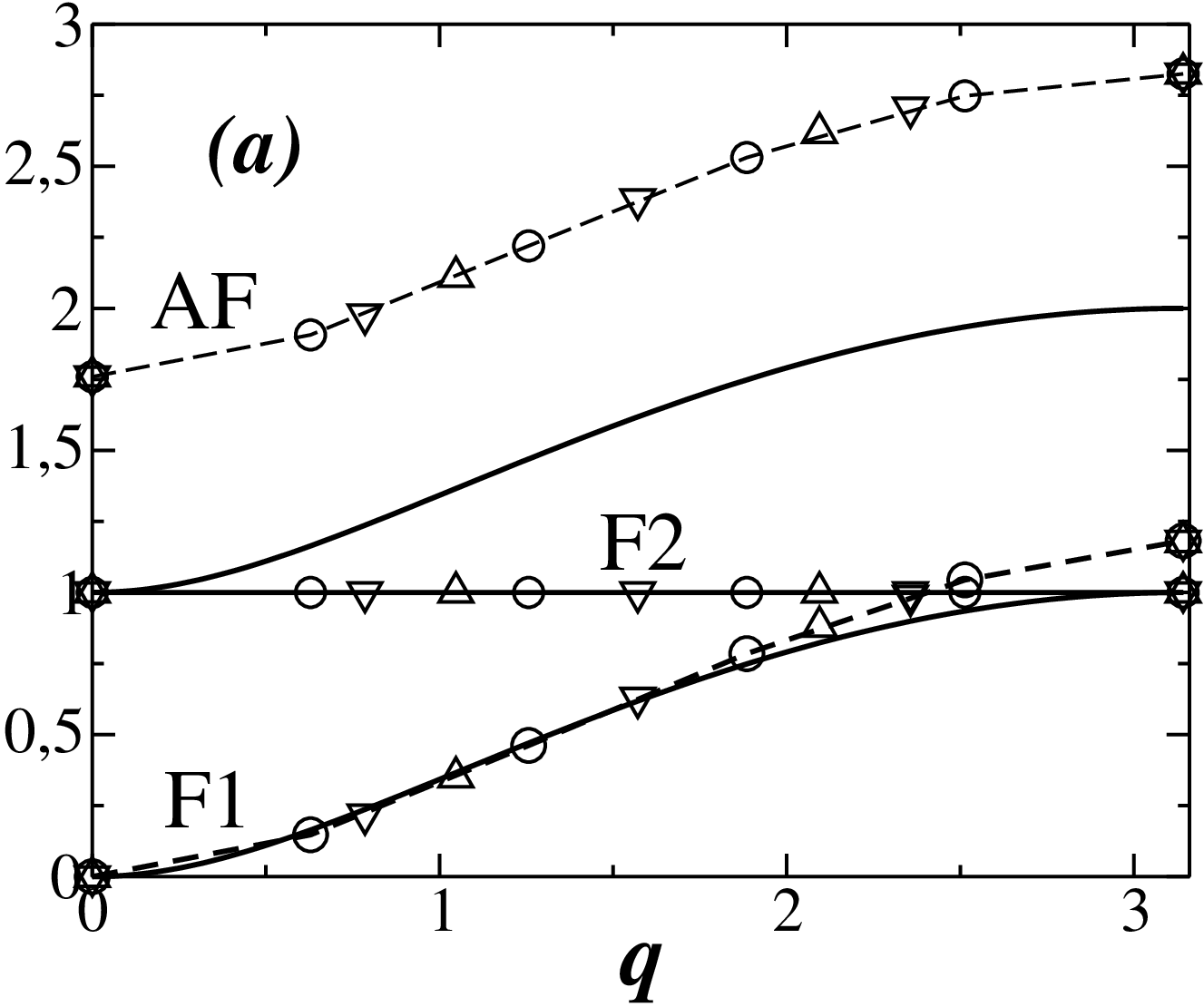}
\includegraphics*[width=6.5cm,clip]{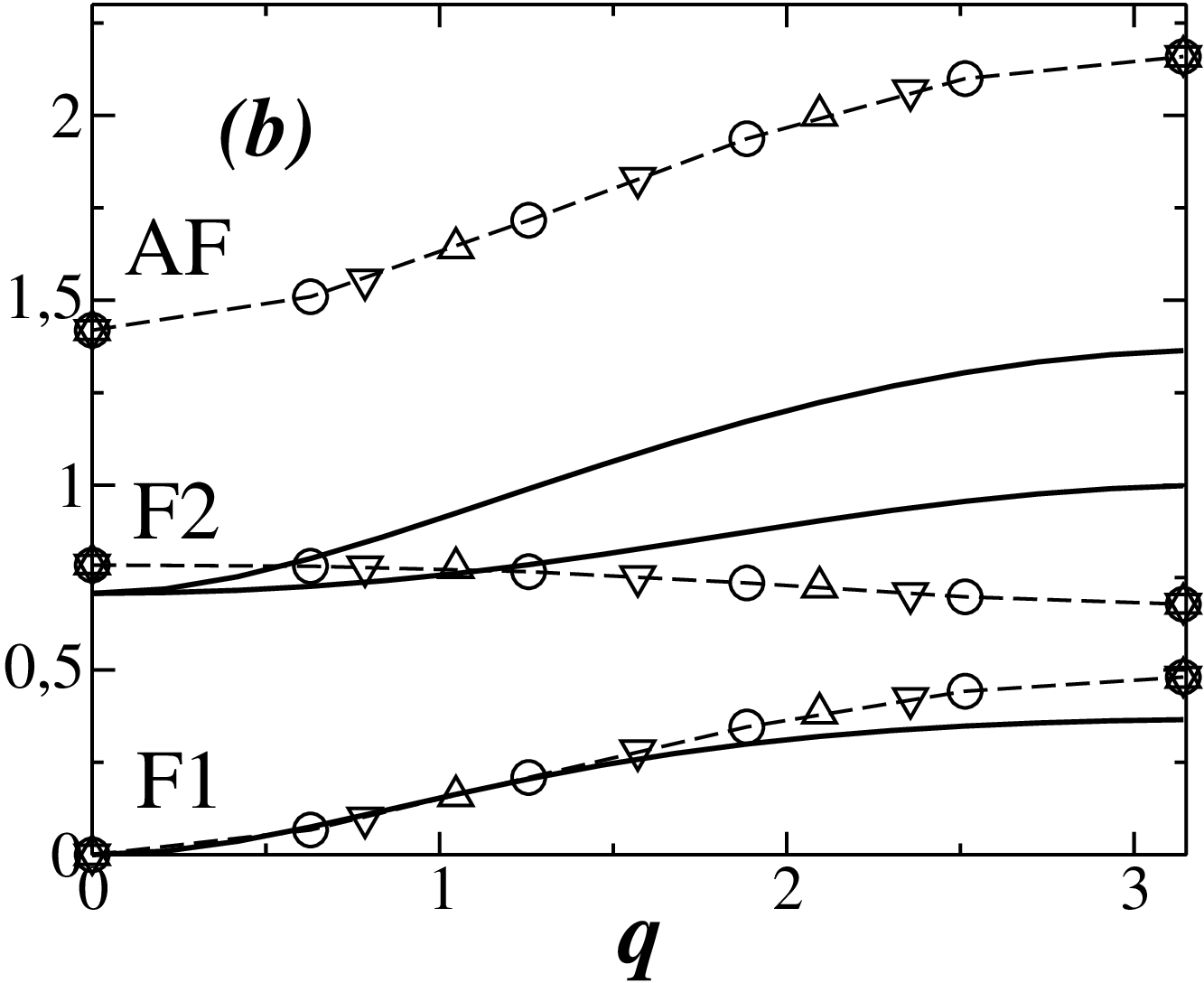}}
\caption{F and AF spin wave modes of the $AB_{2}$ (a) and $ABC$ (b) chains
for $N_c=10$ (circles), 8 (triangles down), 6 (triangles up). Solid lines
are the LSWT results from Ref. \cite{vit1} ($AB_2$) and Ref. \cite{fred} ($ABC$);
dashed lines are guides for the eye.}
\end{center}
\label{enq}
\end{figure}

 Systems with a FERRI GS naturally have F (lowering the GS spin) and
AF (rising the GS spin)
 magnons as their elementary magnetic excitations.
 The two chains have three spin wave branches:
 an AF mode, defined as
 $\Delta E_{S+}(q)=E(S^{z}=S_{g}+1,q)-E_0$;
 and two F ones, derived from
 $\Delta E_{S-}(q)=E(S^{z}=S_{g}-1,q)-E_0$,
 where $E_0$ is the GS energy and $E(S^{z},q)$
 are lowest excitation energies in the sector $\{S^z,q\}$,
 with $q=2\pi l/N_c$ and $l=0,1,...,N_c/2$.
 The acoustical and optical F modes are identified from symmetry
 considerations.
 These modes are depicted in Fig. \ref{enq} for the two chains using the AHM:
 the AF mode has a gap, denoted by $\Delta_{S+}$;
 the F1 mode is gapless, i. e., the Goldstone mode,
 consistent with the symmetry broken
 phase of the chains; the gap of the F2 mode is denoted by $\Delta_{S-}$.

\par For the $AB_{2}$ chain the gapped F2 branch is flat, Fig. \ref{enq}(a),
and the other modes are dispersive.
Since these dispersive modes preserve the local triplet
bond, they are identical to those found in the
spin-$\frac{1}{2}$/spin-$1$ \cite{spin1spin2} chain.
Surprisingly, Linear Spin Wave Theory (LSWT) \cite{vit1} predicts
that $\Delta_{S-}=1$,
very close to our estimated value: $\Delta_{S-}=1.0004J$.
Moreover, a good agreement is found for the gapless F1 branch
in the long wave-length limit.
However,
both LSWT and mean field (MF) theory \cite{vit2} predicts $\Delta_{S+}=1$,
deviating from our estimated exact diagonalization value: $\Delta_{S+}=1.7591J$,
which is in excelent agreement with
numerical and analytical calculations for the
spin-$\frac{1}{2}$/spin-$1$ chain \cite{spin1spin2}.
On the other hand, the
Interacting Spin Wave Theory \cite{yam2}
derives a better result for $\Delta_{S+}$,
but it implies in a higher shift for $\Delta_{S-}$ (flat mode)
not observed in our data of Fig. \ref{enq}(a).
This flat mode indicates
the localized nature of
the excitation, which is associated with the formation of
a singlet state (even parity symmetry) between the $B$ sites in one cell,
while the other cells have odd parity symmetry ($B$ sites in a triplet state).
 Thus, since the symmetries
of the F modes (dispersive and flat) are distinct,
a possible level crossing is not avoided,
as in fact observed in the data of Fig. \ref{enq}(a):
for $q\gtrsim 0.75\pi$ the localized excitation
is below the dispersive one.
\par For the $ABC$ chain we can see in Fig. \ref{enq}(b) that the
three branches are dispersive. Like in the $AB_{2}$ chain,
the AF mode is gapped with $\Delta_{S+}=1.4189$. The LSWT \cite{fred}
 conducts to a qualitative but not quantitative correct
AF branch. However, a good agreement is found for the gapless F1 mode
in the long-wavelength limit. Moreover, the gap for the F2 mode
is estimated to be $\Delta_{S-}=0.6778J$.
We can clearly observe in Fig. \ref{enq}(b)
the level repulsion between the F branches,
which in this case have the same symmetry.
\section{Magnetization Plateaus and Spin Wave Gaps}
\begin{figure}[th]
\begin{center}
\centerline{\includegraphics*[width=6.5cm,clip]{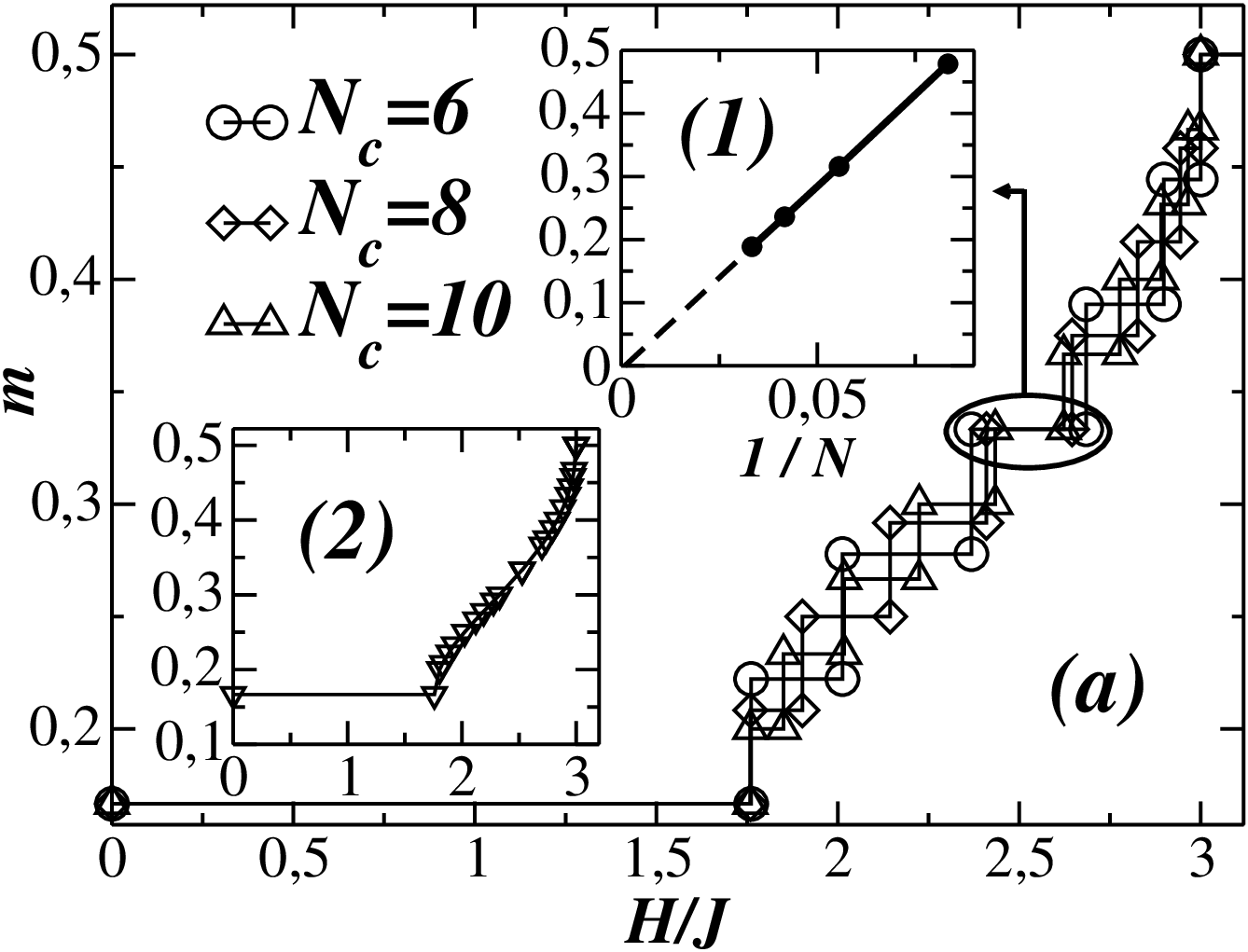}
\includegraphics*[width=6.5cm,clip]{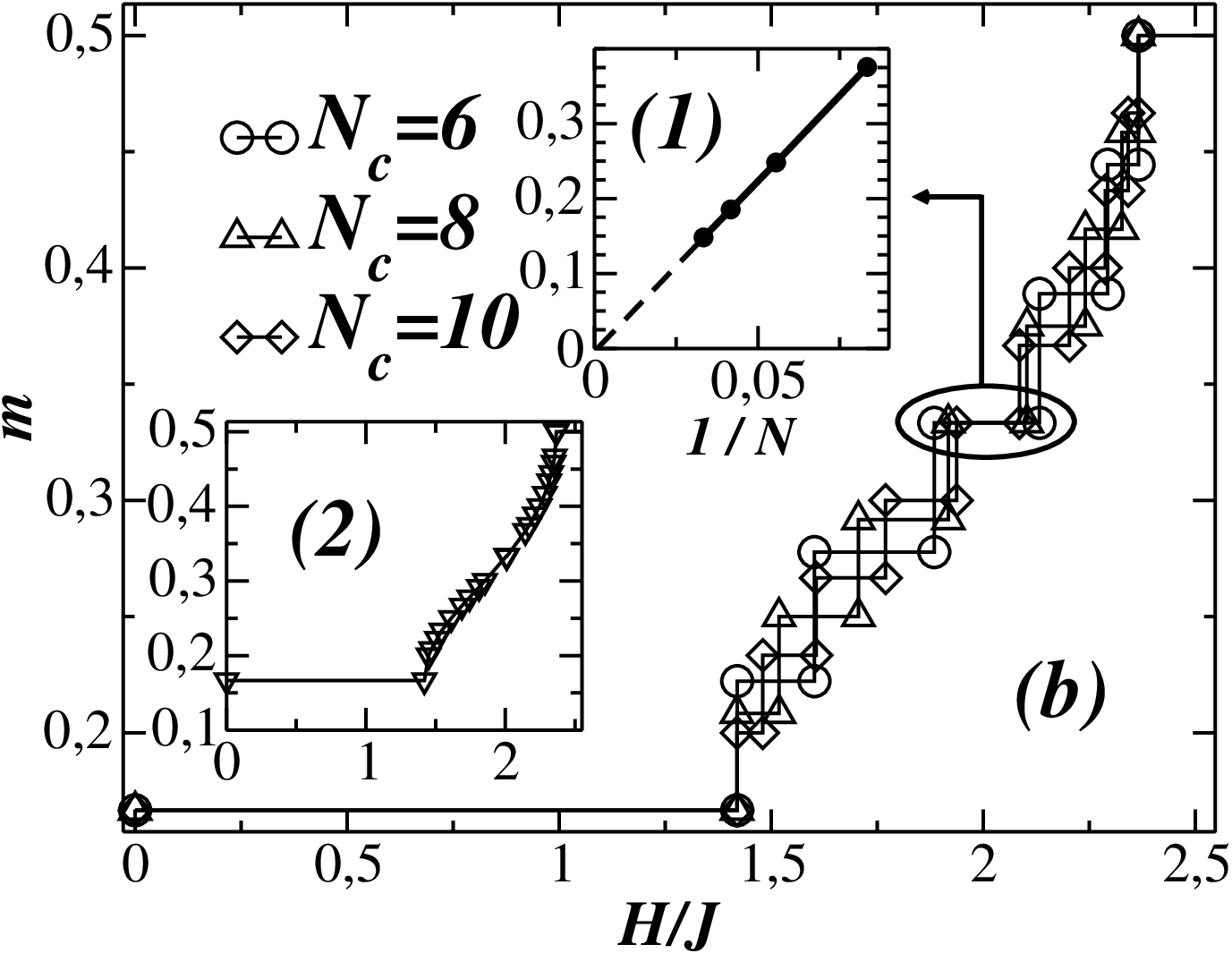}}
\caption{Magnetization as function of the applied magnetic field $H$ for
the $AB_{2}$ (a) and $ABC$ (b) chains using the AHM. The
inset (1) presents a finite size scaling study for $m=1/3$; the
inset (2) is a curve traced from the midpoints of the steps using the
finite size results, except at plateau regions.}
\label{maghei}
\end{center}
\end{figure}

The AF gaps found above are responsible for
plateaus in the magnetization per spin [$m(H)= \langle S^z \rangle/N$] as
function of the applied magnetic field $H$.
In fact, it has been shown \cite{oshikawa}
that if $f(s-m)=\mbox{integer},$
a plateau may appear in the magnetization curve of the AHM.
In the last equation, $s$ is the site total spin, $m$ is the
magnetization and $f$ is the number of sites in one unit
period of the GS for a given value of $H$. The Hamiltonians
that we are considering here have $s=1/2$ and three sites per unit cell,
so, unless the system spontaneously breaks the translation
invariance (e.g., by growing the unit cell), we expect
a plateau at $m=1/6$. This is indeed observed in Fig. \ref{maghei}
and its width matches exactly $\Delta_{S+}$, which measures the
region of stability of the FERRI phase. For higher fields the magnetization
increases in the expected way \cite{oshikawa}, as shown in the inset (2)
of Figs. \ref{maghei} (a) and \ref{maghei} (b), before the magnetization
saturates at $m=1/2$ for $H=3J$ and $H=2.3660J$, respectively. This field
dependent behaviour contrasts with the linear one predicted by MF theory \cite{vit2}.
\par
\begin{figure}[th]
\begin{center}
\centerline{\includegraphics*[width=68mm,clip]{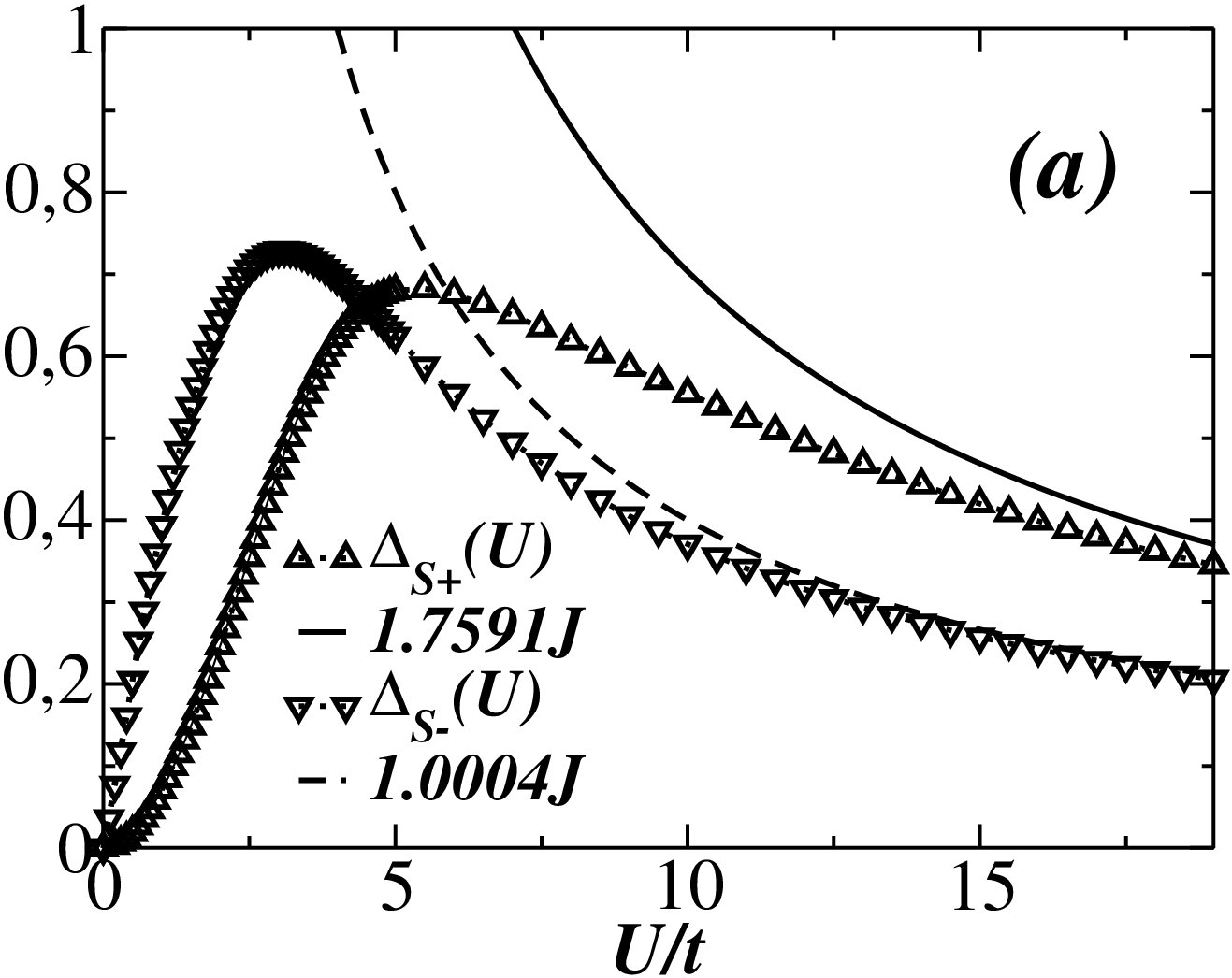}
\includegraphics*[width=65mm,clip]{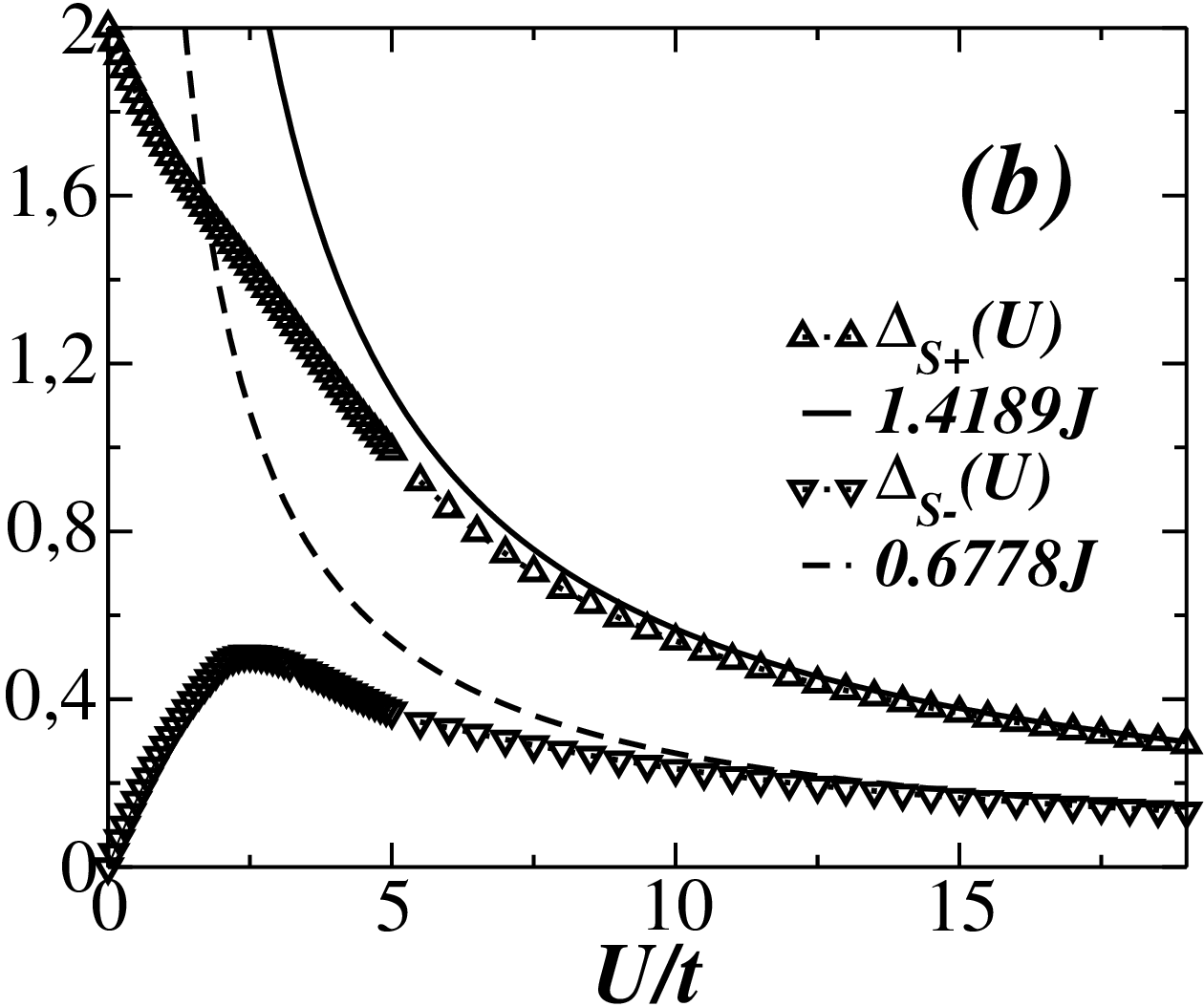}}
\caption{Extrapolated AF ($\Delta_{S+}$) and F2 ($\Delta_{S-}$) spin wave gaps in units of $t$ for the $AB_2$ (a)
and $ABC$ (b) chains as function of $U$ using the HM. Solid (dashed) line is the
AF (F2) gap for the AHM with $J=4t^2/U$; dotted lines are guides for the eye.}
\label{gapsfu}
\end{center}
\end{figure}

 Now we turn to study the
evolution of the AF and the F2 spin wave gaps as function
of $U$ using the HM: $\Delta_{S+}(U)$ (plateau width) and $\Delta_{S-}(U)$,
respectively.
In order to reduce finite size effects in calculating
$\Delta_{S\pm}(U)$ we have used the boundary condition
that minimizes the gap between the GS and excited levels.
Our estimated curves are depicted
in Fig. \ref{gapsfu} using the scaling {\it ansatz} \cite{caprioti} for extrapolation:
$\Delta=\Delta_{\infty}+c_1N^{-2}+c_2N^{-4}$, with $N=9,12,15 \mbox{ and } 18$
for the $AB_2$ chain; $N=6,9,12\mbox{ and }15$ for the $ABC$ chain.
Note that $\Delta_{S+}(U=0)=0$ and $2t$ for the $AB_2$ and $ABC$ chains,
respectively. This can be understood by noticing that
$\Delta_{S+}(U=0)$ is given by the gap between
the two dispersive tight-binding electronic bands.
Moreover, $\Delta_{S+}(U\rightarrow 0)\sim U^x$, $x\simeq 2$,
for the $AB_2$ chain, whereas for the $ABC$ chain $\Delta_{S+}(U\rightarrow 0)$
increases linearly before reaching the tight-binding limit $2t$.
For both chains, we observe a crossover to the Heisenberg
limit: $\Delta_{S+}\sim J$, in agreement with the results of Fig. \ref{enq}.
Further, we notice that $\Delta_{S-}(U\rightarrow 0)$
nullifies linearly with $U$, for both chains,
following the behaviour of the gap between the flat electronic bands \cite{mauricio}.

\section{Conclusions}
In this work we reported a detailed exact diagonalization study
of the half-filled Hubbard model on bipartite quasi-one-dimensional
lattices. These chains have experimental motivation in inorganic
and organic ferrimagnetic polymers. We have focused on the description of
their ferrimagnetic ground state properties and magnetic excitations.
In particular, we have revealed the non-trivial physical behaviour
exhibited by these chains as function of the Coulomb coupling strength.

\section{Acknowledgements}

We acknowledge several useful discussion with A. L. Malvezzi.
This work was supported by CNPq, Finep, FACEPE and CAPES (Brazilian agencies).




\end{document}